\documentclass[11pt]{article}
\usepackage{acl-hlt2011}
\usepackage{times}
\usepackage{latexsym}
\usepackage{amsmath}
\usepackage{multirow}
\usepackage{url}
\usepackage{color}
\usepackage{graphicx}

\title{Image Mining from Gel Diagrams in Biomedical Publications}

\author{
  Tobias Kuhn \\
  Department of Pathology \\
  Yale University School of Medicine \\
  New Haven, CT, USA \\
  {\tt kuhntobias@gmail.com} \\\And
  Michael Krauthammer \\
  Department of Pathology \\
  Yale University School of Medicine \\
  New Haven, CT, USA \\
  {\tt michael.krauthammer@yale.edu} \\
}

\date{}

\begin{document}
\maketitle
\begin{abstract}
Authors of biomedical publications often use gel images to report experimental results such as protein-protein interactions or protein expressions under different conditions. Gel images offer a way to concisely communicate such findings, not all of which need to be explicitly discussed in the article text. This fact together with the abundance of gel images and their shared common patterns makes them prime candidates for image mining endeavors. We introduce an approach for the detection of gel images, and present an automatic workflow to analyze them. We are able to detect gel segments and panels at high accuracy, and present first results for the identification of gene names in these images. While we cannot provide a complete solution at this point, we present evidence that this kind of image mining is feasible.
\end{abstract}

\section{Introduction}

A recent trend in the area of literature mining is the inclusion of images in the form of figures from biomedical publications \cite{yu2006bioinform,zweigenbaum2007briefbioinform,peng2008bioinform}. This development benefits from the fact that an increasing number of scientific articles are published as open access publications. This means that not just the abstracts but the complete texts including images are available for data analysis. Among other things, this enabled the development of query engines for biomedical images like the Yale Image Finder \cite{xu2008bioinform} and the BioText Search Engine \cite{hearst2007bioinform}.

Gel images are a very frequent type of image in the biomedical literature. They are the result of gel electrophoresis, which is a common method to analyze DNA, RNA and proteins. Southern, Western and Northern blotting \cite{southern1975jmolbiol,alwine1977nas,burnette1981analbiochem} are among the most common applications of gel electrophoresis. The resulting experimental artifacts are often shown in biomedical publications in the form of gel images as evidence for the discussed findings such as protein-protein interactions or protein expressions under different conditions. According to our experience, about 15\% of all subfigures (i.e. independent parts of a figure) are gel images. Often, not all details of the results shown in these images are explicitly stated in the caption or the article text. For these reasons, it would be of high value to be able to reliably mine the relations encoded in these images.

\begin{figure*}[t]
\begin{center}
\footnotesize
\begin{tabular}{rl}
\begin{tabular}{c}
\includegraphics[scale=2.2]{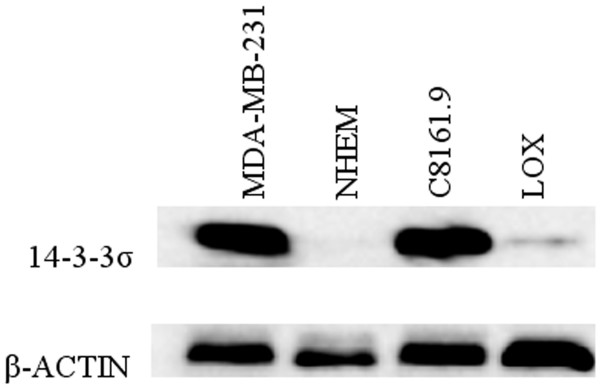}
\end{tabular}
&
\begin{tabular}{|l|l|l|}
\hline
\textbf{Condition} & \textbf{Measurement} & \textbf{Result} \\
\hline
MDA-MB-231 & 14-3-3$\sigma$ & high expression \\
\hline
NHEM & 14-3-3$\sigma$ & no expression \\
\hline
C8161.9 & 14-3-3$\sigma$ & high expression \\
\hline
LOX & 14-3-3$\sigma$ & low expression \\
\hline
MDA-MB-231 & $\beta$-actin & high expression \\
\hline
NHEM & $\beta$-actin & high expression \\
\hline
C8161.9 & $\beta$-actin & high expression \\
\hline
LOX & $\beta$-actin & high expression \\
\hline
\end{tabular}
\\
\bigskip\\
\begin{tabular}{c}
\includegraphics[scale=0.6]{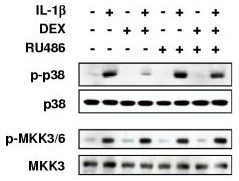}
\end{tabular}
&
\begin{tabular}{|lll|l|l|}
\hline
\multicolumn{3}{|l|}{\textbf{Condition}} & \textbf{Measurement} & \textbf{Result} \\
\hline
IL-1$\beta$ (--) & DEX (--) & RU486 (--) & p-p38 & low expression \\
\hline
IL-1$\beta$ (+) & DEX (--) & RU486 (--) & p-p38 & high expression \\
\hline
IL-1$\beta$ (--) & DEX (+) & RU486 (--) & p-p38 & no expression \\
\hline
IL-1$\beta$ (+) & DEX (+) & RU486 (--) & p-p38 & low expression \\
\hline
IL-1$\beta$ (--) & DEX (--) & RU486 (+) & p-p38 & no expression \\
\hline
IL-1$\beta$ (+) & DEX (--) & RU486 (+) & p-p38 & high expression \\
\hline
IL-1$\beta$ (--) & DEX (+) & RU486 (+) & p-p38 & low expression \\
\hline
IL-1$\beta$ (+) & DEX (+) & RU486 (+) & p-p38 & high expression \\
\hline
... & & & ... & ... \\
\hline
\end{tabular}
\\
\end{tabular}
\caption{Two examples of gel images from biomedical publications (PMID 19473536 and 15125785) with tables showing the relations that could be extracted from them}
\label{fig:example}
\end{center}
\end{figure*}

A closer look at gel images reveals that they follow regular patterns to encode their semantic relations. Figure \ref{fig:example} shows two typical examples of gel images together with a table representation of the involved relations. The ultimate objective of our approach (for which we can only present a partial solution here) is to automatically extract at least some of these relations from the respective images, possibly in conjunction with classical text mining techniques. The first example shows a Western blot for detecting two proteins (14-3-3$\sigma$ and $\beta$-actin as a control) in four different cell lines (MDA-MB-231, NHEM, C8161.9, and LOX, the first of which is used as a control). There are two rectangular gel segments arranged in a way to form a $2\times4$ grid for the individual eight measurements combining each protein with each cell line. A gel diagram can be considered a kind of matrix with pictures of experimental artifacts as content. The tables to the right illustrate the semantic relations encoded in the gel diagrams. Each relation instance consists of a condition, a measurement and a result. The proteins are the entities being measured under the conditions of the different cell lines. The result is a certain degree of expression indicated by the darkness of the spots (or brightness in the case of white-on-black gels). The second example is a slightly more complex one. Several proteins are tested against each other in a way that involves more than two dimensions. In this case, the use of ``+'' and ``--'' labels is a frequent technique to denote the different possible combinations of a number of conditions. Apart from that, the principles are the same. In this case, however, the number of relations is much larger. Only the first eight of overall 32 relation instances are shown in the table to the right. In such cases, the text rarely mentions all these relations in an explicit way, and the image is therefore the only accessible source.

\section{Background}

In principle, image mining involves the same processes as classical literature mining \cite{debruijn2002ijmi}: document categorization, named entity tagging, fact extraction, and collection-wide analysis. However, there are some subtle differences. Document categorization corresponds to image categorization, which is different in the sense that it has to deal with features based on the two-dimensional space of pixels, but otherwise the same principles of automatic categorization apply. Named entity tagging is different in two ways: pinpointing the mention of an entity is more difficult with images (a large number of pixels versus a couple of characters), and OCR errors have to be considered. Fact extraction in classical literature mining involves the analysis of the syntactic structure of the sentences. In images, in contrast, there are rarely complete sentences, but the semantics is rather encoded by graphical means. Thus, instead of parsing sentences, one has to analyze graphical elements and their relation to each other. The last process, collection-wide analysis, is a higher-level problem, and therefore no fundamental differences can be expected. Thus, image mining builds upon the same general stages as classical text mining, but with some subtle but important differences.

Image mining on biomedical publications is not a new idea. It has been applied for the extraction of subcellular location information \cite{murphy2004ksce}, the detection of panels of fluorescence microscopy images \cite{qian2008bioinform}, the extraction of pathway information from diagrams \cite{kozhenkov2012bioinform}, and the detection of axis diagrams \cite{kuhn2012amia}. Also, there is a large amount of existing work on how to process gel images \cite{lemkin1997electrophoresis,luhn2003proteomics,cutler2003proteomics,rogers2003proteomics,zerr2005nar} and databases have been proposed to store the results of gel analyses \cite{schlamp2008gene}. These techniques, however, take as input plain gel images, which are not readily accessible from biomedical papers, because they make up just parts of the figures. Furthermore, these tools are designed for researchers who want to analyze their gel images and not to read gel diagrams that have already been analyzed and annotated by a researcher. Therefore, these approaches do not tackle the problem of recognizing and analyzing the labels of gel images. Some attempts to classify biomedical images include gel figures \cite{rodriguez2009bioinform}, which is, however, just the first step in locating them and analyzing their labels and their structure. To our knowledge, nobody has yet tried to perform image mining on gel diagrams.

\section{Approach and Methods}

\begin{figure*}[t]
\begin{center}
\includegraphics[width=\textwidth]{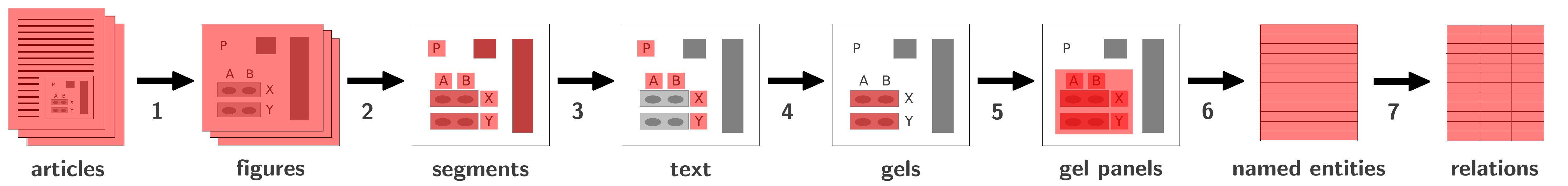}
\caption{The procedure of our approach: (1) figure extraction, (2) segmentation, (3) text recognition, (4) gel detection, (5) gel panel detection, (6) named entity recognition, and (7) relation extraction.}
\label{fig:procedure}
\end{center}
\end{figure*}

Figure \ref{fig:procedure} shows the procedure of our approach to image mining from gel diagrams. It consists of seven steps: figure extraction, segmentation, text recognition, gel detection, gel panel detection, named entity recognition and relation extraction.

Using structured article representations, the first step is trivial. For the steps two and three, we rely on existing work. The focus of this paper lies on steps four, five and six: the detection of gels and gel panels and the recognition of named entities. We sketch how step seven could be implemented, but we cannot provide a solution at this point.

To practically evaluate our approach, we ran our pipeline on the entire open access subset of PubMed Central (though not all figures made it through the whole pipeline due to technical difficulties).

\subsection{Figure Extraction}

A large portion of the articles of the open access subset of the PubMed Central database are available as structured XML files with additional image files for the figures. We only use these articles so far, which makes the figure extraction task very easy. It would be more difficult, though definitely feasible, to extract the figures from PDF files or even bitmaps of scanned articles.

\subsection{Segmentation and Text Recognition}

For the next two steps --- segment detection and subsequent text recognition ---, we rely on our previous work \cite{xu2010jbi,xu2011bsec}. This method includes the detection of layout elements, edge detection, and text recognition with a novel pivoting approach. For optical character recognition (OCR), the Microsoft Document Imaging package is used, which is available as part of Microsoft Office 2003. Overall, this approach has been shown to perform better than other existing approaches for the images found in biomedical publications \cite{xu2010jbi}. We do not go into the details here, as this paper focuses on the subsequent steps.

Due to some limitations of the segmentation algorithm when it comes to rectangles with low internal contrast (like gels), we applied a complementary very simple rectangle detection algorithm.

\subsection{Gel Segment Detection}

Based on the results of the above-mentioned steps, we try to identify gel segments. Such gel segments typically have rectangular shapes with darker spots on a light gray background, or --- less commonly --- white spots on a dark background. We decided to use machine learning techniques to generate classifiers to detect such gel segments. To do so, we defined 39 numerical features for image segments: the coordinates of the relative position (within the image), the relative and absolute width and height, 16 grayscale histogram features, three color features (for red, green and blue), 13 texture features based on \newcite{haralick1973tsmc}, and the number of recognized characters.

To train the classifiers, we took a random sample of 500 figures, for which we manually annotated the gel segments. In the same way, we obtained a second sample of another 500 figures for testing the classifiers. We used the Weka toolkit and opted for random forest classifiers based on 75 random trees.\footnote{We also tried other types of classifiers including support vector machines, but we achieved the best results with random forests.} Using different thresholds to adjust the trade-off between precision and recall, we generated a classifier with good precision and another one with good recall. Both of them are used in the next step.

\subsection{Gel Panel Detection}

A gel panel typically consists of several gel segments and comes with labels describing the involved genes, proteins, and conditions. For our goal, it is not sufficient to just detect the figures that contain gel panels, but we also have to extract their positions within the figures and to access their labels. This is not a simple classification task, and therefore machine learning techniques do not apply that easily. For that reason, we used a detection procedure based on hand-coded rules.

In a first step, we group gel segments to find contiguous gel regions that form the center part of gel panels. To do so, we start with looking for segments that our high-precision classifier detects as gel segments. Then, we repeatedly look for adjacent gel segments, this time applying the high-recall classifier, and merge them. Two segments are considered neighbors if they are at most 50 pixels apart\footnote{We are using absolute distance values at this point. A more refined algorithm could apply some sort of relative measure. However, the resolution of the images does not vary that much, which is why absolute values worked out well so far.} and do not have any text segment between them. Thus, segments which could be gel segments according to the high-recall classifier make it into a gel panel only if there is at least one high-precision segment in their group. The goal is to detect panels with high precision, but also to detect the complete panels and not just parts of them. In the given situation, precision is more important than recall, because low recall can be leveraged by the large number of available gel images.

As a next step, we collect the labels in the form of text segments located around the detected gel regions. For a text segment to be attributed to a certain gel panel, its nearest edge must be at most 30 pixels away from the border of the gel region and its farthest edge must not be more than 150 pixels away. We end up with a representation of a gel panel consisting of two parts: a center region containing a number of gel segments and a set of labels in the form of text segments located around the center region.

To evaluate this algorithm, we collected yet another sample of 500 figures. For these, we manually checked whether the algorithm is able to detect the presence and the (approximate) position of the gel panels.

\subsection{Named Entity Recognition}

The next step is to recognize the named entities mentioned in the gel labels. To this aim, we investigated whether we are able to extract the names of genes and proteins from gel diagrams.\footnote{Apart from genes and proteins, we plan to include the names of cell lines and drugs in future work.} To do so, we tokenized the label texts and looked for entries in the Entrez Gene database to match the tokens. This look-up is done in a case-sensitive way, because many names in gel labels are acronyms, where the specific capitalization pattern can be critical to identify the respective entity. We excluded tokens that have less than three characters, are numbers (Arabic or Latin), or correspond to common short words (retrieved from a list of the 100 most frequent words in biomedical articles). In addition, we extended this exclusion list with 22 general words that are frequently used in the context of gel diagrams, some of which coincide with gene names according to Entrez.\footnote{These words are: \emph{min}, \emph{hrs}, \emph{line}, \emph{type}, \emph{protein}, \emph{DNA}, \emph{RNA}, \emph{mRNA}, \emph{membrane}, \emph{gel}, \emph{fold}, \emph{fragment}, \emph{antigen}, \emph{enzyme}, \emph{kinase}, \emph{cleavage}, \emph{factor}, \emph{blot}, \emph{pro}, \emph{pre}, \emph{peptide}, and \emph{cell}.}

Since gel electrophoresis is a method to analyze genes and proteins, we would expect to find more such mentions in gel labels than in other text segments of a figure. By measuring this, we get an idea of whether the approach works out or not. In addition, we manually checked the gene and protein names extracted from gel labels after running our pipeline on 2000 random figures.

\subsection{Relation Extraction}

For the last step, relation extraction, we cannot present concrete results at this point. After recognizing the named entities, we would have to disambiguate them, identify their semantic roles (condition, measurement or something else), align the gel images with the labels, and ultimately quantify the degree of expression. To improve the quality of the results, combinations with classical text mining techniques should be considered. This is all future work. We expect to be able to profit to a large extent from existing work to disambiguate protein and gene names \cite{rinaldi2008genomebiol,tanabe2002bioinform} and to detect and analyze gel spots (see the existing work mentioned above).

\section{Results}

Table \ref{tab:geldetection} shows the result of the gel detection classifier. We generated three different classifiers from the training data, one for each of the threshold values 0.15, 0.3 and 0.6. Lower threshold values lead to higher recall at the cost of precision, and vice versa. In the balanced case, we achieved an F-score of 75\%. To get classifiers with precision or recall over 90\%, F-score goes down significantly, but stays in a sensible range. These two classifiers (thresholds 0.15 and 0.6) are used in the next step. To interpret these values, one has to consider that gel segments are greatly outnumbered by non-gel segments. Concretely, only about 3\% are gel segments. Accuracy measures take this into account. The accuracy of the presented classifiers, measured as the area under the ROC curve, is 98.0\%.\footnote{This measure includes all thresholds from 0 to 1.}
\begin{table}[t]
\begin{center}
\begin{tabular}{r|rrr}
Threshold & Precision & Recall & F-score \\
\hline
0.15 & 0.439 & 0.909 & 0.592 \\
0.30 & 0.765 & 0.739 & 0.752 \\
0.60 & 0.926 & 0.301 & 0.455 \\
\end{tabular}
\caption{The results of the gel segment detection classifiers}
\label{tab:geldetection}
\end{center}
\end{table}

The results of the gel panel detection algorithm are shown in Table \ref{tab:paneldetection}. The precision is 95\% at a recall of 38\%, leading to an F-score of 54\%.
\begin{table}[t]
\begin{center}
\begin{tabular}{rrr}
Precision & Recall & F-score \\
\hline
0.951 & 0.379 & 0.542 \\
\end{tabular}
\caption{The results of the gel panel detection algorithm}
\label{tab:paneldetection}
\end{center}
\end{table}

Table \ref{tab:pipeline} shows the results of running the pipeline on PubMed Central. We started with about 410\,000 articles, the entire open access subset of PubMed Central at the time we downloaded them (February 2012). We successfully parsed the XML files of 94\% of these articles (for the remaining articles, the XML file was missing or not well-formed, or other unexpected errors occurred). The successful articles contained around 1\,100\,000 figures, for some of which our segment detection step encountered image formatting errors or other internal errors, or was just not able to detect any segments. We ended up with more than 880\,000 figures, in which we detected about 86\,000 gel panels, i.e. roughly ten out of 100 figures. For each of them, we found on average 3.6 labels with recognized text. After tokenization, we identified about 76\,000 gene names in these gel labels, which corresponds to 6.8\% of the tokens. Considering all text segments (including but not restricted to gel labels), only 3.3\% of the tokens are detected as gene names.\footnote{The low numbers are partially due to the fact that a considerable part of the tokens are ``junk tokens'' produced by the OCR step when trying to recognize characters in segments that do not contain text.}
\begin{table}[t]
\begin{center}
\begin{tabular}{l|r}
Total articles & 410\,950 \\
Processed articles & 386\,428 \\
\hline
Total figures from processed articles & 1\,110\,643 \\
Processed figures & 884\,152 \\
\hline
Detected gel panels & 85\,942 \\
Detected gel panels per figure & 0.097 \\
Detected gel labels & 309\,340 \\
Detected gel labels per panel & 3.599 \\
\hline
Detected gene tokens & 1\,854\,609 \\
Detected gene tokens in gel labels & 75\,610 \\
Gene token ratio & 0.033 \\
Gene token ratio in gel labels & 0.068 \\
\end{tabular}
\caption{The results of running the pipeline on the open access subset of PubMed Central}
\label{tab:pipeline}
\end{center}
\end{table}
% Figures processed: 884152
% Gel panels detected: 85942
% Gel labels detected: 309340
% Other labels detected: 9421924
% Gene tokens in gel labels: 75610/1119250 = 0.06755417
% Gene tokens in other labels: 1778999/54399681 = 0.03270238

Table \ref{tab:geneprecision} shows the results of the evaluation of the detection algorithm for gene and protein names. Almost two-thirds of the detected gene/protein tokens (65.3\%) were correctly identified. 9\% thereof were correct but could be more specific, e.g. when only ``actin'' was recognized for ``$\beta$-actin''. The incorrect cases (34.6\%) can be split into two classes of roughly the same size: some recognized tokens were actually not mentioned in the figure but emerged from OCR errors; other tokens were correctly recognized but incorrectly classified as gene or protein references.
\begin{table}[t]
\begin{center}
\begin{tabular}{l|rr}
& absolute & relative \\
\hline
\textbf{Total} & \textbf{156} & \textbf{100.0\%} \\
\hline
\textbf{Incorrect} & \textbf{54} & \textbf{34.6\%} \\
-- Not mentioned (OCR & 28 & 17.9\% \\
{\color{white}--} errors) & & \\
-- Not references to genes & 26 & 16.7\% \\
{\color{white}--} or proteins & & \\
\hline
\textbf{Correct} & \textbf{102} & \textbf{65.3\%} \\
-- Partially correct (could & 14 & 9.0\% \\
{\color{white}--} be more specific) & & \\
-- Fully correct & 88 & 56.4\% \\
\end{tabular}
\caption{Number of recognized gene/protein tokens in 2000 random figures}
\label{tab:geneprecision}
\end{center}
\end{table}
% Total recognized gene/protein tokens: 156
% Incorrectly recognized tokens: 28
% Correct tokens but no genes/proteins: 26
% Correct gene/protein tokens that could be more specific: 14
% Perfectly recognized gene/protein tokens: 88

\section{Discussion}

The presented results show that we are able to detect gel segments with high accuracy, which allows us to subsequently detect whole gel panels at a high precision. The recall of the panel detection step is relatively low, but with about 38\% still in a reasonable range. As mentioned above, we can leverage the high number of available figures, which makes precision more important than recall.

Running our pipeline on the whole set of open access articles from PubMed Central, we were able to retrieve 85\,942
potential gel panels (around 95\% of which we can expect to be correctly detected). The detection of gene and protein names reveals that they are more than twice as frequent in gel labels than in other text segments, which is consistent with what one would expect. This simple gene detection step performs reasonably well with a precision of about 65\%, though there is certainly room for improvement.

It seems reasonable to assume that these results can be combined with existing techniques of term disambiguation and gel spot detection at a satisfactory level of accuracy. We plan to investigate this in future work.

Our results indicate that it is feasible to extract relations from gel images, but it is clear that this procedure is far from perfect. The automatic analysis of bitmap images seems to be the only efficient way to extract such relations from existing publications, but other publishing techniques should be considered for the future. The use of vector graphics instead of bitmaps would already greatly improve any subsequent attempts of automatic analysis. A further improvement would be to establish accepted standards for different types of biomedical diagrams in the spirit of the Unified Modeling Language, a graphical language widely applied in software engineering since the 1990s. Ideally, the resulting images could directly include semantic relations in a formal notation, which would make relation mining a trivial procedure. If authors are supported by good tools to draw diagrams like gel images, this approach could turn out to be feasible even in the near future.

\section{Conclusions}

Successful image mining from gel diagrams in biomedical publications would unlock a large amount of valuable data. Our results show that gel panels and their labels can be detected with high accuracy, applying machine learning techniques and hand-coded rules. We also showed that genes and proteins can be detected in the gel labels with satisfactory precision.

Based on these results, we believe that this kind of image mining is a promising and viable approach to provide more powerful query interfaces for researchers, to gather relations such as protein-protein interactions, and to generally complement existing text mining approaches. At the same time, we believe that an effort towards standardization of scientific diagrams such as gel images would greatly improve the efficiency and precision of image mining at relatively low additional costs at the time of publication.

\section*{Acknowledgments}

This study has been supported by the National Library of Medicine grant 5R01LM009956.

\bibliographystyle{acl}
\bibliography{gels}

\end{document}